# The TRANSGUIDE: Ultra-bright directional light emission from any refractive index material

**Abstract:**

This report introduces the Transfer Waveguide (TRANSGUIDE); an ultra-thin flat technology that promises light emitting applications a practical solution to total internal reflection light trapping and diverging emission. By invoking reciprocity, light can be temporarily stored in the form of a virtual-dipole and recovered back again.

Hossam Galal

[Galal Space](#)

hgalal4716@gmail.com

**Introduction**

Light extraction and directional emission, especially from high-refractive index materials, is a bottleneck for many light emission applications such as Light Emitting Diodes (LEDs), lasers, and Single Photon Sources (SPSs). A significant amount of light is lost inside the structure of the light emitting device, whether due to Total Internal Reflection (TIR) light trapping or due to the presence of metallic elements. The scheme theoretically outlined in this text promises light extraction efficiencies approaching 100% in conjunction with highly directional beam profiles, and is applicable to different wavelength and material systems.

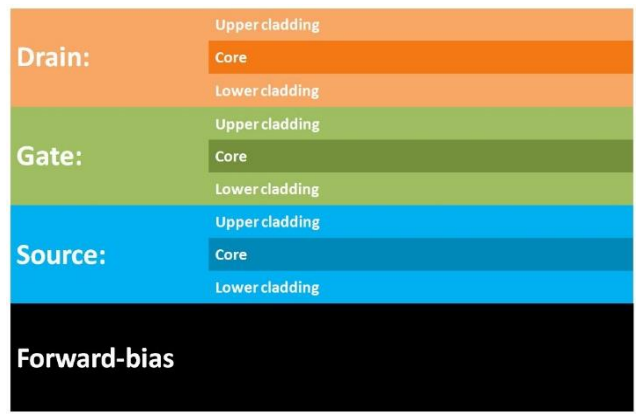

**Fig. 1 Scheme outline.** Flat waveguides packed on a reflecting layer, forming a three-stage stack (cross-section view).

Waveguides have long been known for guiding light along their axes. Here, they are used differently, and it shall be shown that they hold as much promises from their sides. Three simple plain waveguides: the Source; the Gate; and the Drain are packed one upon another, on the Forward-bias layer, to form a transistor-like architecture referred to as the Transfer Waveguide (TRANSGUIDE), schematically shown in Fig. 1.

Essentially, light is emitted within the Source waveguide, from emitters with their dipole orientation parallel to the Gate waveguide's axis. The Forward-bias is a reflecting layer that directs the emitted photons towards the Drain waveguide. The overall structure is designed to support a photon flow from the Source to the Drain, i.e. along the tunnel.

The following discussions are highlights on how a three-stage stack, with waveguide building blocks constructed from very simple materials; can bring light sources, whether classical or quantum, to another level. In laying down the concepts, the Silicon Vacancy (SiV) color center in diamond is considered as the light emitting platform [1]. This type of quantum emitter has recently emerged as a promising SPS candidate at 738 nm; however, TIR due to diamond's high-refractive index remains an issue to be addressed. As simple as it gets; by putting plain oxide layers together, for example, a practical solution to this longstanding problem will be tailored in the upcoming examples. The proposed layered structure is free of any advanced nanostructures or engineered metamaterials, and offers a broad technological tolerance margin. The refractive indices of $TiO_2$, ZnO, $Al_2O_3$, $SiO_2$, diamond, and silver at the aforementioned wavelength read: 2.54 [2], 1.97 [3], 1.76 [4], 1.45 [5], 2.4 [6], and 0.033+5.1i [7], respectively. Choice of materials is solely for outlining the working concept and is not restricted. Outperforming results are conceived with selective material combinations.

**Leakage tunneling**

**Imaginary-Source**

To this extent, as far as understood, any of the three waveguides is typically comprised of a core and a cladding, with the cladding consisting of a lower part and an upper part. With reference to the configuration shown in Fig. 2a, the Source's core and cladding are defined from the same material; meaning that, there is no refractive index contrast between them. The Source, in this case, is considered imaginary since its core and cladding are merged together; this implies that imaginary physical interfaces define the boundaries between them. Presumably, the emitters are located in close proximity to the center of the Source, marked with the origin of the coordinate system. The Gate functions in a similar manner to an Insulator Metal Insulator (IMI) waveguide. In this arrangement, the Gate is merged; meaning that the Gate's lower cladding is merged with the Source's upper cladding, i.e. same material. On the other side of the Gate, a typical dielectric waveguide serves as the Drain. A real physical interface draws the boundary between the Gate and the Drain, via the refractive index contrast between their contiguous claddings. Dimensions of the individual waveguides can be read in the caption of Fig. 2.

Tolerances, in terms of the Light Extraction Efficiency (LEE) and the Radiation Half Angle (RHA), are analyzed along the z-coordinate up to ±20 nm emitter position displacement ($\Delta d$) from the origin, shown in Fig. 2b. Lateral displacements, along the x and y coordinates, are completely irrelevant owing to the structure's translational symmetry. The figure of merit LEE is defined as: the Power Radiated from the Drain along the z-coordinate in the forward direction ($P_{rad}$) divided by the Total Power emitted by the dipole within the Source ($P_{tot}$). The radiation pattern is quantified by the RHA measured at half-maximum. Whether moving towards the Gate (+$\Delta d$) or towards the Forward-bias (-$\Delta d$), the LEE remains in the vicinity of 78% with a 27° RHA single-lobe pattern. Reminding that, the background medium is air, i.e. free-space emission. Not shown in Fig. 2b, the Purcell factor takes values between 1 and 2.

The modes supported in this arrangement are plotted in Fig. 2c. Radiative modes are identified with real effective refractive index values $Re(n_{eff})$ = [0:1], while non-radiative modes feature a $Re(n_{eff})$ > 1 (shaded area in Fig. 2c). In this context, non-radiative modes can be thought of as loss channels. The behaviors of these non-radiative modes are understood from the successive field profiles shown in Fig. 2d-g. In Fig. 2d, due to the finite thickness of the diamond layer, the two Surface Plasmon Polariton (SPP) modes at each of its interfaces merge together within the diamond layer, giving rise to a hybrid guided mode in the Source. SPP is considered an interesting phenomenon; however, for light emitting devices it can be bothersome. The second mode, in Fig. 2e, is guided in the Gate. This mode is partially converted into a TIR contribution along the $TiO_2$-$SiO_2$ transition. A similar TIR contribution could also be seen earlier in Fig. 2d. The two remaining modes in Fig. 2f,g are guided in the Drain: the first is a fundamental mode concentrated in the core, while the second is a higher order mode in the cladding. The cladding mode is capped with a TIR contribution along the SiO$_2$-air transition. The tunneling from the Source to the Drain's cladding, under the gating effect, can be clearly remarked in Fig. 2g.

**Real-Source**

A noticeable difference between this arrangement and the previous one is the physical structure of the Source. Here in Fig. 3a, there is a refractive index contrast between the Source's core and cladding; implying that, real physical interfaces define the boundaries between them. As in the previous example, the emitters are initially positioned close to the center of the Source. The Gate is again merged; however, this time the Gate's lower cladding is SiO$_2$ instead of diamond.

Let us take a look at the tolerance plot shown in Fig. 3b, where the robustness of the modified Source structure is investigated. Apparently, the figures of merit LEE and RHA show higher emitter position insensitivity close to 90% and 19°, respectively. The radiated power is funneled into a symmetric single-lobe, with Purcell factor values between 1 and 2 (not shown).

Overall, from Fig. 3c-g; the hybrid Source guided mode becomes leakier, stepping down. With the electric field concentrated in the Source's cladding, more light is transferrable from the Source to the Drain.

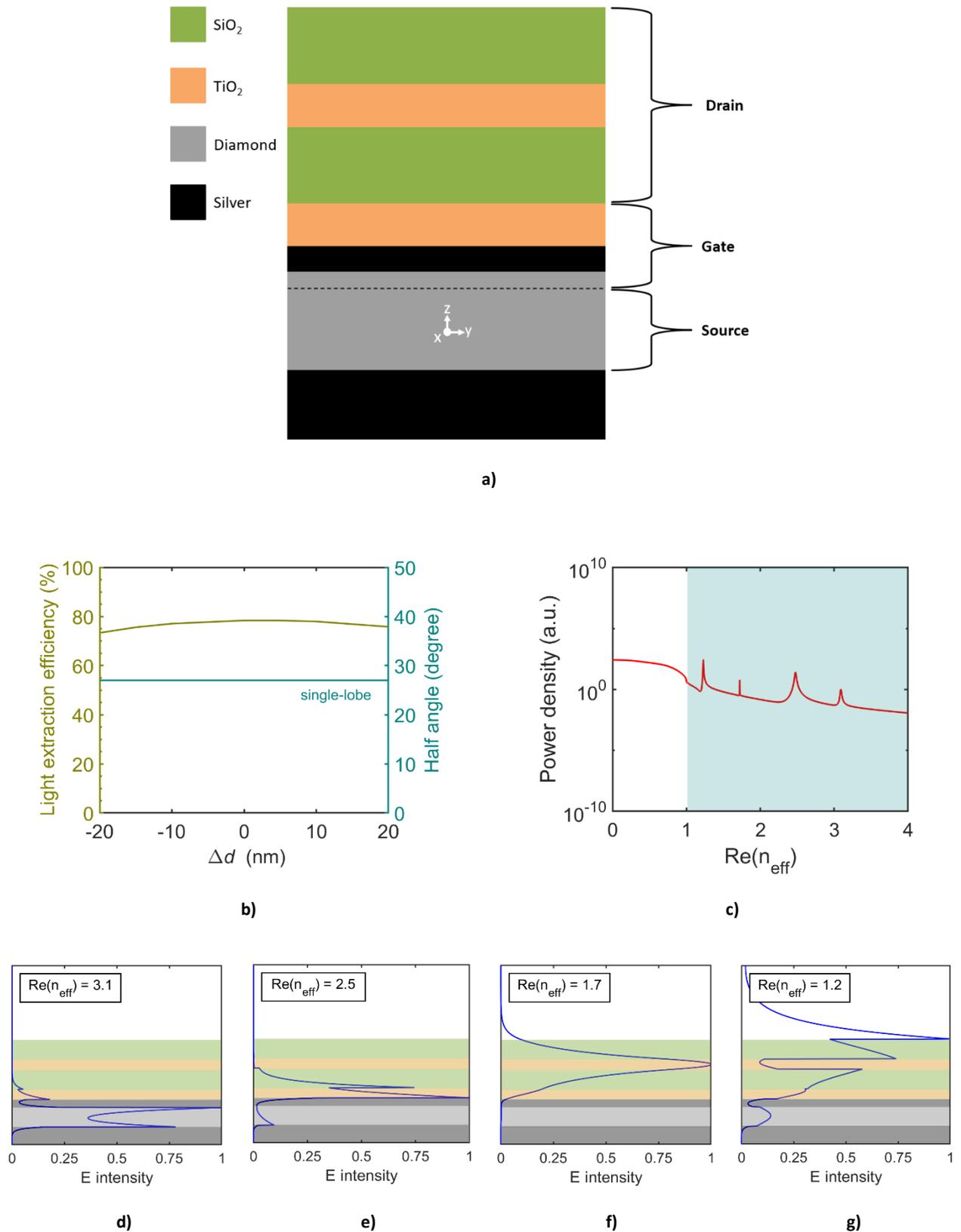

**Fig. 2 Imaginary-Source merged-Gate configuration (free-space emission).** a) Stack dimensions: [Forward-bias: (100 nm)], [Source: (25 nm, 50 nm, 25 nm)], [Gate: (10 nm, 45 nm, 58 nm)], [Drain: (112 nm, 58 nm, 112 nm)]. Coordinate system origin marks the center of the Source. b) Figures of merit as a function of the emitter position displacement from the center of the Source. c) Power density, on a logarithmic scale, as a function of the real effective refractive index. d)-g) Normalized modal field distributions of the four non-radiative modes shaded in c). Color mapping is in correspondence with a).

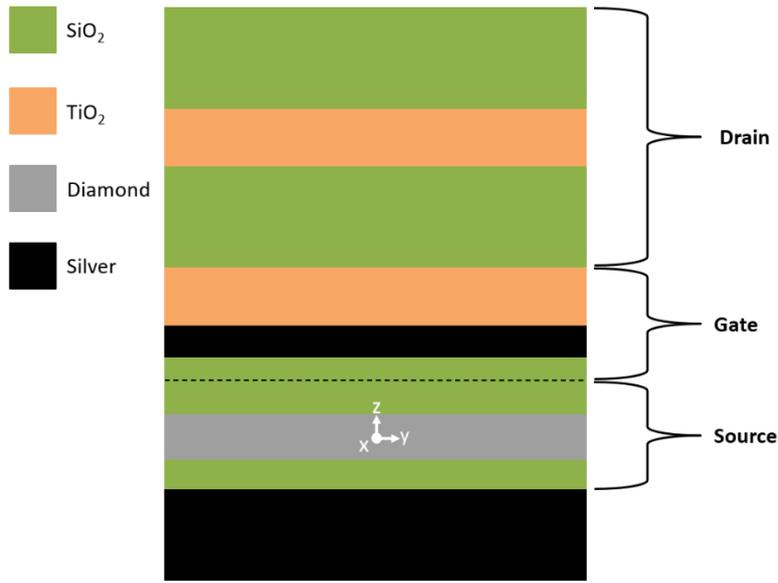

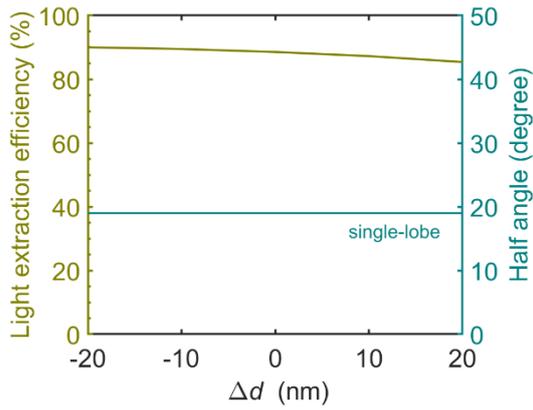
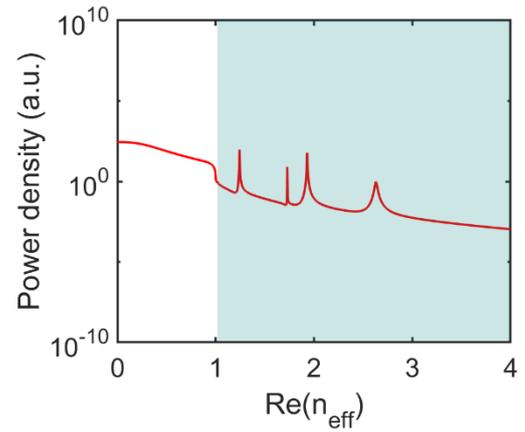

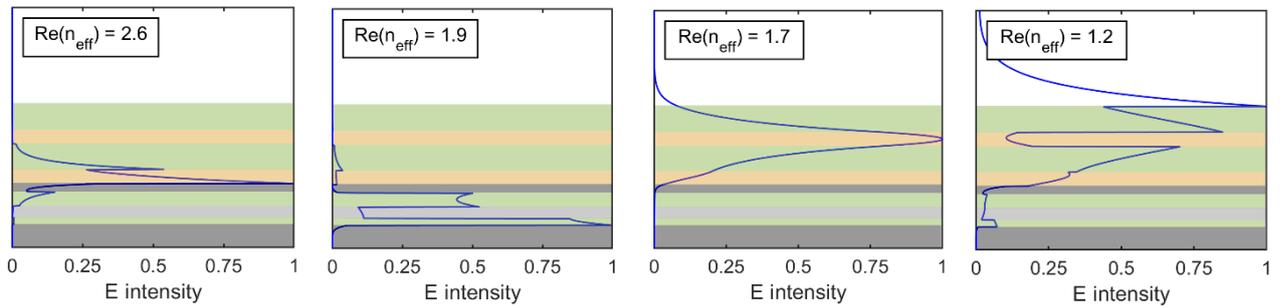

**Fig. 3 Real-Source merged-Gate configuration (free-space emission).** a) Stack dimensions: [Forward-bias: (100 nm)], [Source: (30 nm, 50 nm, 30 nm)], [Gate: (31 nm, 35 nm, 62 nm)], [Drain: (112 nm, 58 nm, 112 nm)]. Coordinate system origin marks the center of the Source. b) Figures of merit as a function of the emitter position displacement from the center of the Source. c) Power density, on a logarithmic scale, as a function of the real effective refractive index. d)-g) Normalized modal field distributions of the four non-radiative modes shaded in c). Color mapping is in correspondence with a).

**Pinching**

Now that we have identified the dominant loss channels, herein we take a look at the main radiant modes. At first, consider the imaginary-Source merged-Gate configuration's radiant mode plotted in Fig. 4a. The Source's field tunnels its way through the Gate, and ends up stored in the Drain's cladding. Notice that the mode is pinched twice: once in the core of the Gate, and a second in the core of the Drain. A pinch is technically a tightening of the light's path; in the same sense as, a pinch for the charge carriers in a Field Effect Transistor (FET). The Drain has been configured not to hold on to the mode for long, and eventually the light bursts into free-space.

A similar scenario takes place in Fig. 4b, for the real-Source merged-Gate configuration; however, this time the Source's field is heavily depleted, with an additional pinch in the Source's core. The higher LEE delivered by this configuration, as compared to the former, translates from the enhanced mode tunneling and pinching.

The corresponding electric field cross-sectional mappings shown in Fig. 4c,d illustrate the light transfer along the tunnel, and the eventual burst into free-space. For the latter configuration, the Drain appears fuller while the Source is almost completely drained. Obviously, the burst is seeded from the Drain, and is pinched in the three waveguide cores. Even though the emitters are incubated in the Source, their light emission rather builds up in the Drain. The light radiated from the Drain, visualized in the far-field, is shown directly beneath in Fig. 4e,f; with the left-panels for the examples under consideration, and the right-panels for selective designs featuring different pinching effects. The idea here is to show that pinching is capable of routing and reshaping the emission beam profile. The RHA can readily be reduced, for example, from 27° to 15°, in Fig. 4e; and from 19° to 12.5°, in Fig. 4f. Tighter angles are realizable.

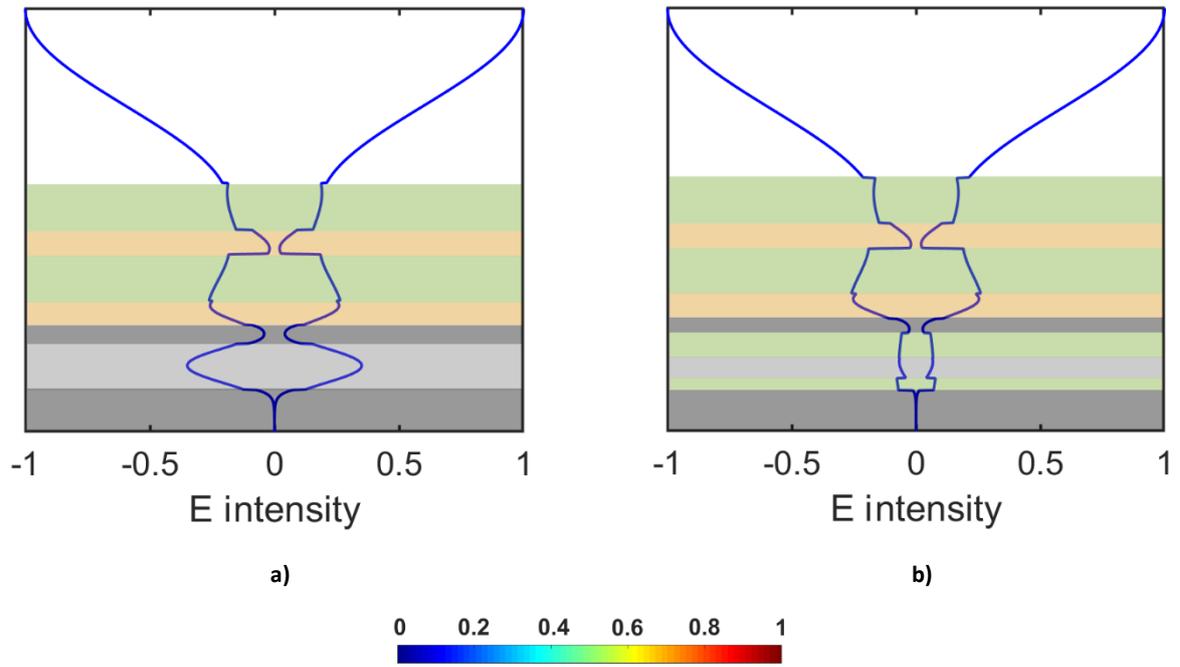

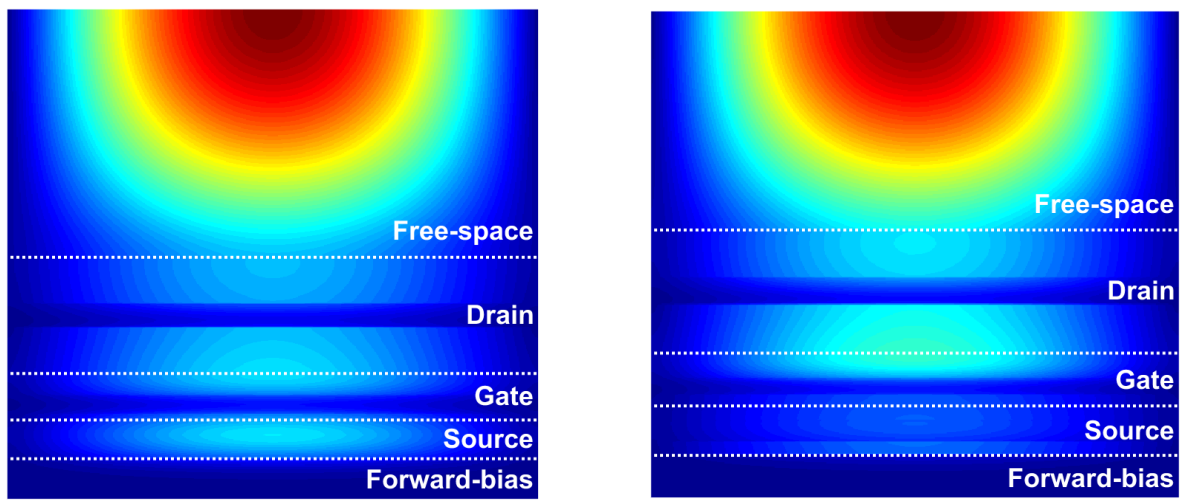

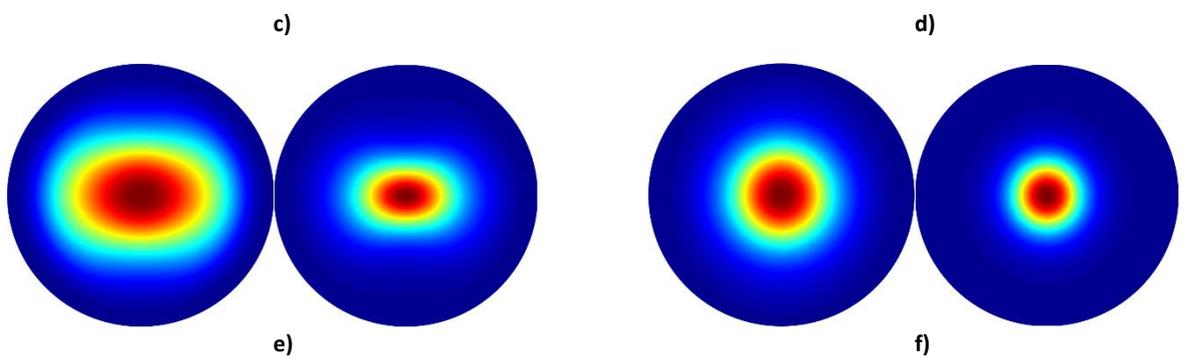

**Fig. 4 Radiated power (free-space emission): imaginary-Source merged-Gate vs real-Source merged-Gate.** a), b) The main radiant mode for the configurations with an imaginary-Source merged-Gate and a real-Source merged-Gate, respectively. Color mapping is in correspondence with Fig. 2a and Fig. 3a, respectively. c), d) Normalized electric field magnitude cross-sections for a) and b), respectively. e) (left) Normalized far-field radiation pattern for a); (right) for a configuration similar to a) but featuring a different pinching factor. f) (left) Normalized far-field radiation pattern for b); (right) for a configuration similar to b) but featuring a different pinching factor. All scales are normalized.

**Gate trenching**

One of the Gate's functionalities is to regulate the light transfer between the Source and the Drain. In the two previous examples, the Gate was merged with the Source. Here in Fig. 5, we will consider something different: physically detaching the Source from the Gate. This is realized by using a different material for the Gate's lower cladding; where in Fig. 5a the diamond lower cladding is replaced with ZnO for the imaginary-Source configuration, and in Fig. 5b the SiO$_2$ lower cladding is replaced with Al$_2$O$_3$ for the real-Source configuration. The overall refractive index profiles of the structures are depicted in Fig. 5c,d respectively. For both configurations, the Gate's upper cladding remains TiO$_2$. The impact of the Gate trenching is addressed in Fig. 5e,f; with the earlier merged-Gate plots recalled from Fig. 2b and Fig. 3b, respectively. The free-space LEE approaches 81% for the imaginary-Source trenched-Gate configuration; and 91% for the real-Source trenched-Gate configuration, exhibiting a flatter plateau response.

In this context, two arguments might arise. First; the enhancement in the LEE delivered by switching from an imaginary-Source (Fig. 2a) to a real-Source (Fig. 3a), might be misinterpreted as a result of the suppression of SPPs in response to the inclusion of low-refractive index layers, i.e. replacing the silver-diamond interfaces in the former case with silver-SiO$_2$ interfaces in the latter case. The LEE plots in Fig. 5f prove this argument wrong. This is understood from the fact that, Gate trenching the real-Source configuration is equivalent to replacing the silver-SiO$_2$ interface with a silver-Al$_2$O$_3$ interface; bearing in mind that Al$_2$O$_3$ has a higher refractive index than SiO$_2$. Even when the dipole emitter approaches the Gate, i.e. ($+\Delta d$), the LEE for the trenched-Gate case increases with respect to the merged-Gate case. One would have expected it to rather decrease [8].

The other argument that might come into mind; Gate trenching the imaginary-Source configuration of Fig. 2a with ZnO in Fig. 5a, can be visualized as replacing the silver-diamond interface in the former with a silver-ZnO interface in the latter. Since the refractive index of ZnO is lower than that for diamond, it makes sense that the LEE is higher for the ZnO trenched case; owing to the less coupling to SPPs [8]. However; this perception is misleading, because, trenching the Gate with an even lower refractive index material, such as $SiO_2$, deteriorates the LEE rather than enhancing it.

From here, I would like to emphasize on the fact that, the TRANSGUIDE's potentials are higher when the individual waveguides are well defined. It has got nothing to do with the suppression of SPPs by the inclusion of low-refractive index layers; one can already see that the effective refractive index values are high enough to provoke strong SPPs. After all, the three waveguides have been stitched together in a way, such that all competing loss channels are settled down.

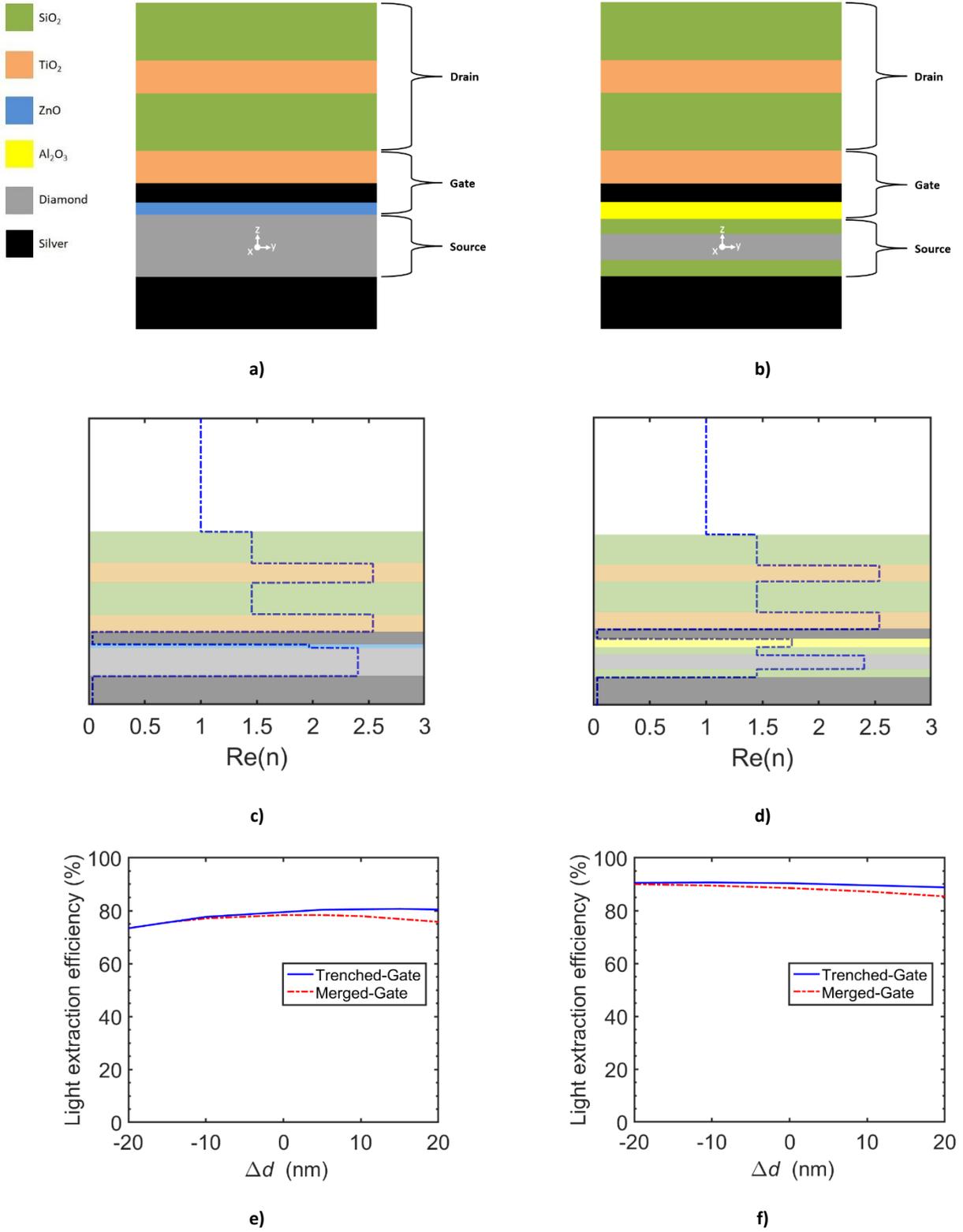

**Fig. 5 Gate trenching (free-space emission).** a) Imaginary-Source: the Gate's lower cladding and the Source's upper cladding are different materials. Configuration dimensions: [Forward-bias: (100 nm)], [Source: (25 nm, 50 nm, 25 nm)], [Gate: (12 nm, 45 nm, 62 nm)], [Drain: (112 nm, 68 nm, 112 nm)]. b) Real-Source: the Gate's lower cladding and the Source's upper cladding are different materials. Configuration dimensions: [Forward-bias: (100 nm)], [Source: (30 nm, 50 nm, 30 nm)], [Gate: (29 nm, 35 nm, 62 nm)], [Drain: (112 nm, 58 nm, 112 nm)]. c), d) Refractive index profiles of the configurations in a) and b), respectively. Layer color mapping is in correspondence with a) and b). e), f) Comparisons between the trenched-Gate configurations and the merged-Gate configurations (plots recalled from Fig. 2b and Fig. 3b) in terms of the LEE as a function of the emitter position displacement from the center of the Source, for the configurations in a) and b).

**State of the art**

The trend in the past years has been to incorporate nanostructures or engineered materials in light emitting devices, taking advantage of the latest advances in nanofabrication. It turns out that; this just makes the whole process more complex and adds up additional bills. Even the reported free-space LEEs have not really benefited from that sophisticated technology; in fact, image blurring and diffraction in illumination applications [9] and bandwidth modulation are a symptom. All this comes at the expense of the radiation pattern, requiring additional external-integrated optics to account for the large radiation angle spans and the distorted beam profile. In practice, the spatial positioning of the dipole light emitters with respect to the incorporated nanostructures becomes very critical; thereby, further tightening the technological tolerance margin.

In a different direction, there is that claimed planar antenna scheme, reported first in [10]. Let us take that approach on a one-to-one comparison with the TRANSGUIDE. The planar antenna, when applied to a 100 nm diamond active medium, struggles with a 40% free-space LEE [8]. One step further, the advanced version of the said antenna, applied to a 50 nm diamond active medium, the free-space LEE saturates close to 80% [8]. When it comes to the TRANSGUIDE; we have already witnessed free-space LEEs near 81% on that 100 nm diamond active medium, and 91% on that 50 nm diamond active medium. This superiority comes in response to stitching two symmetric dielectric waveguides together by means of an antisymmetric plasmonic waveguide.

Technically speaking, the scheme in [8] principally relies on the suppression of SPPs with the inclusion of very low-refractive index layers, ideally $n \approx 1$. Such stringent requirement makes the scheme highly restrictive and inapplicable to free-space emission applications; especially with the fact that typical charge-transport material layers feature a relatively high-refractive index. Meanwhile, the TRANSGUIDE has no problems with SPPs, and is comfortable with high-refractive index materials.

**Medium vs free-space emission**

At this stage, the free-space LEE exceeds 90% with the promoted leakiness of the Source. The power contributions of the four non-radiative modes (Fig. 2c-g and Fig. 3c-g) are negligible compared to the radiant continuum. That having been said, if these modes were to be plotted on a linear scale, most of them would even be hardly seen. However; out of the four, the Drain mode $Re(n_{eff}) = 1.2$ amounts for the largest share, with a significant amount of the modal power seized in the Drain's cladding. With reference to Fig. 2g and Fig. 3g one can see that, in the latter case, more light builds up in the Drain's cladding. For both cases, the mode is capped with a TIR contribution.

Interfacing the Drain's outermost cladding with a material, having a relatively higher refractive index; sets the aforementioned Drain mode free. This is demonstrated in Fig. 6a,b, with the blue plots representing the real-Source trenched-Gate configuration of Fig. 5b packaged in epoxy or silicone gel (n=1.5 −1.55); while the red dashed plots are for free-space background. The shaded area in Fig. 6a marks the non-radiative modes for the packaging plot. As a result, the Drain's cladding mode is completely converted into a radiant mode, and the Drain's remaining guided core mode is further weakened; giving rise to a 92% LEE, at least for this configuration. The origin of this 1−5% incrementation is depicted in Fig. 6b. The Source in turn is evacuated and more light is transferred along the tunnel, with an exponential field rise in the Drain's outermost cladding. At the packaging's interface, TIR no longer exists, and the mode bursts out of the Drain. The LEE, in packaging and in free-space, as a function of the emitter position displacement ($\Delta d$) is plotted in Fig. 6c,d; for the imaginary-Source and for the real-Source both trenched-Gate configurations, respectively. To mention, however, the LEE for the inferior imaginary-Source trenched-Gate configuration approaches 86% in packaging.

We are now left with three non-radiative modes, one of which has already been strongly weakened. With the careful tailoring of the materials and design, we can further weaken or even get rid of the leftover non-radiative modes.

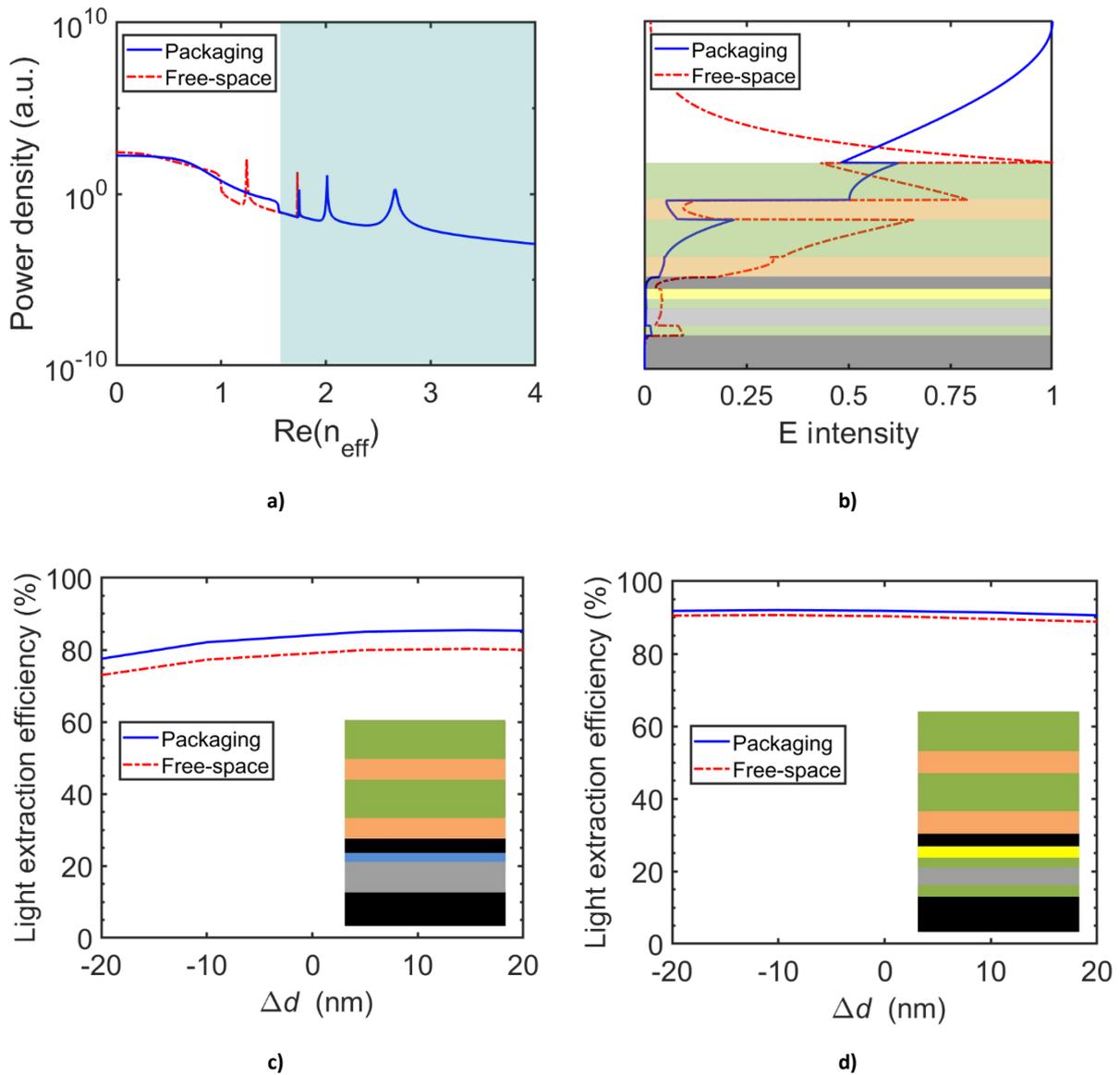

**Fig. 6 Packaging (blue) vs free-space (red) emission.** a) Power densities as a function of the real effective refractive index, for the real-Source trenched-Gate configuration. Shaded area marks the non-radiative modes for the packaging plot. b) Transformation of the Drain's guided cladding mode, for the real-Source trenched-Gate configuration. Normalized scale. c), d) The LEE, packaging vs free-space; as a function of the emitter position displacement from the center of the Source for the imaginary-Source trenched-Gate and the real-Source trenched-Gate configurations, respectively.

Before we move on to the next section, let me come back to the earlier SPP argument and wrap it up with the two real-Source trenched-Gate packaged examples shown in Fig. 7. The cyan plot is a ZnO trench, while the yellow plot is the $Al_2O_3$ trench recalled from Fig. 6d. Even though ZnO has a higher refractive index than $Al_2O_3$, the ZnO trench LEE picks up a ~3% and hits a plateau near 95%.

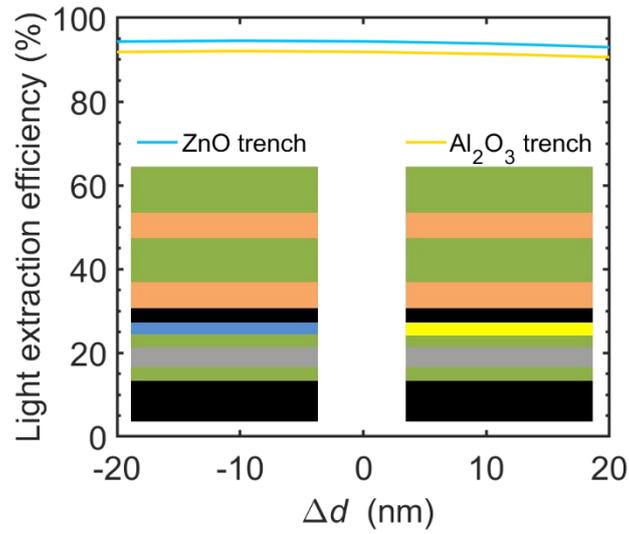

**Fig. 7 Real-Source Gate trenches (packaging emission).** The LEE as a function of the emitter position displacement from the center of the Source. The two plots are real-Source trenched-Gate configurations, with the yellow $Al_2O_3$ plot recalled from Fig. 6d.

**Reciprocity**

Light detection applications, or to say fluorescence and energy conversion systems, are more sensitive to the light's electric field component. Nanostructures have been explored for this purpose, with focus on mediating light-matter interactions through enhancing the electric field strength. Polarization plays an important role, where ideally the incoming light's electric field has to maintain a certain alignment with respect to the nanostructure. Typically, electric field augmentation is spatially limited to superficial nanometers-size "hotspots" around sharp edges or within gaps [11]. Most literature studies in this regard assume an ideal scenario, where the incoming light beam is a well-defined polarization plane wave. In real-life, the beam can arrive with any permissible polarization state, with a finite spot-size that is unnecessarily uniformly distributed; this is a common practice, for example, in LiDAR and sensing applications. Non-uniform exposure can lead to dramatic outcomes and unexpected suppressions [12]; especially when the nanostructure is arranged in a periodic array, exploiting collective interference effects [11].

In what follows, and by invoking reciprocity, we take a look into what the TRANSGUIDE offers in this regard.

**Beam exposure**

An efficient light-matter interface would be one that: first, does not waste the incoming light beam by being specific for certain polarization components while transparent for others; second, is independent of the spot-size and the profile of the incoming light beam. To address these issues, a randomly polarized Gaussian profile will be considered for exposure. The numerical aperture of a Gaussian beam is a very critical parameter, and directly relates to the beam profile specifications: spot-size, amplitude distribution, divergence angle, etc.

Now that we have agreed to a realistic exposure, recall the real-Source trenched-Gate design that we came across earlier in Fig. 5b. The structure is exposed to the incoming Gaussian beam head-on, over free-space, as shown in Fig. 8a. For the sake of highlighting the TRANSGUIDE's reciprocity, assume that probing the SiV color center under resonant-excitation is of interest; regardless of whether or not it is physically feasible. The 738 nm wavelength Gaussian beam pump is shone on the Drain, with the SiV presumably located at the center of the Source waveguide, i.e. the origin.

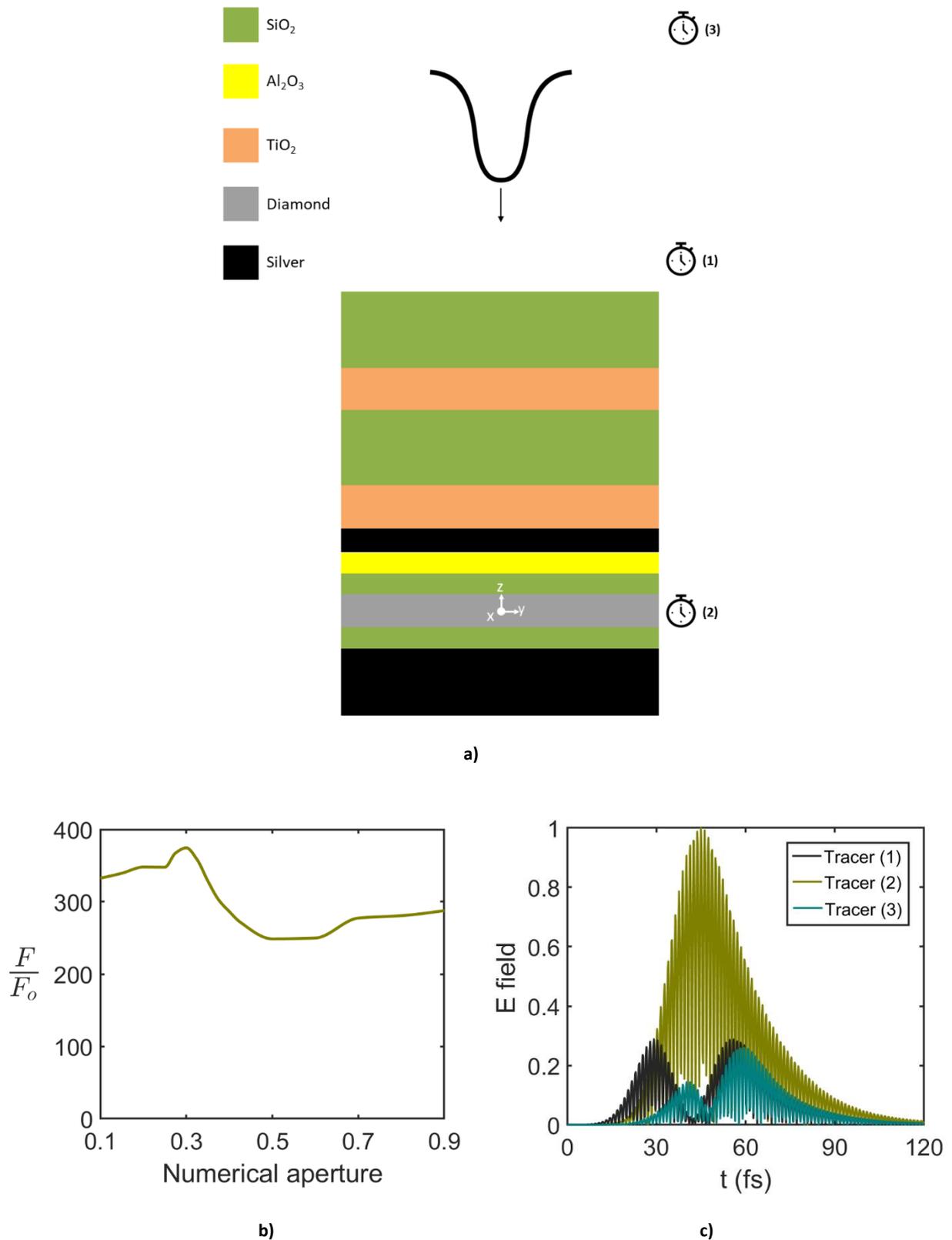

**Fig. 8 Free-space Gaussian beam exposure:** a) A real-Source trenched-Gate configuration exposed to a 738 nm wavelength arbitrary polarization Gaussian beam, along the (-ve) z-axis. The structure's design parameters can be read in the caption of Fig. 5b. The coordinate system marks the center of the Source waveguide, where ideally the xy-focal plane is located. b) The enhancement factor computed over the focal plane in response to the beam's numerical aperture values. c) Round-trip response for the 0.3 numerical aperture incident Gaussian beam. Tracer locations are pointed out in a).

The electric field is observed over an xy-plane, known as the focal plane, located at the center of the Source waveguide. For this argument, we will recall the "normalized electric energy density" factor [13,14]. The reference scenario would be when the incoming beam interacts with the diamond active medium standalone, i.e. the grey slab in Fig. 8a; and in this case we can define a corresponding reference normalized electric energy density factor ($F_o$). The other scenario would be when the incoming beam interacts with the complete layered structure shown in Fig. 8a, giving rise to a corresponding normalized electric energy density factor ($F$). The enhancement factor is of interest, quantified by ($F/F_o$).

As the beam's numerical aperture varies, the enhancement factor easily maintains values above 200, shown in Fig. 8b. This response applies to any beam polarization, since the TRANSGUIDE structure is rotationally symmetric around the z-axis. The peak in the vicinity of ~0.3 numerical aperture emphasizes on the scheme's reciprocity. This can be understood by referring back to the structure's principle design, where the outgoing light beam in essence was meant to be emitted from the Drain along the (+ve) z-axis with a RHA on the order of 19° (comparable to ~0.3 numerical aperture). With the incoming beam following the same trajectory of the outgoing beam, the structure's maximum potential can be reproduced. The Gate, in practice, is a two-way port. This enhancement is not locally limited to the focal plane; in fact, it extends over the entire volume of the Source waveguide's core, in agreement with what perceived earlier. The broadband nature of the scheme extends its applicability beyond resonant-excitation applications (see for example Fig. 9c).

### Virtual-dipole

In order to have a better understanding of the waveguides' responses to the incoming beam, maybe we should consider tracing the beam at different time frames. Carrying out this analysis for a 0.3 numerical aperture would probably be ideal for observing the structure's reciprocity. Three beam tracers are placed at the different locations, pointed out to in Fig. 8a: tracer (1) along the z-axis between the Gaussian beam aperture and the TRANSGUIDE, tracer (2) at the center of the Source waveguide, and tracer (3) along the z-axis behind the Gaussian beam aperture. The picked-up signals are plotted on a normalized scale in Fig. 8c.

The plot from tracer (1) is the event onset, where the impinging pulse is registered as the first packet around 30 fs. While this packet continues its way propagating along the (-ve) z-axis, it is partially back-reflected of the Gate; and accordingly, the reflected portion propagates backwards along the (+ve) z-axis and is registered around 40 fs on tracer (3). The portion of the beam that experienced no reflection ends up in the Source's core. This portion is significantly enhanced and is the only packet registered by tracer (2), around 45 fs. Since diamond is a transparent material, and in this particular study we have not defined any dipolar system to absorb the incoming light; the packet is released back to where it came from. On its return journey, it is first registered around 56 fs on tracer (1) and later on around 60 fs on tracer (3). Notice that the packet's form resembles a dipolar-exponential-decay lifetime envelope, and is preserved at all three registered instances. This packet is a signature of the TRANSGUIDE's reciprocity. Back-reflected light, more or less, has the same symmetric envelope as the incident light pulse, but with a weaker amplitude.

A final point-out on Fig. 8c. The earlier discussions have repeatedly stressed on the necessity that the dipole emitters maintain an in-plane polarization within the Source, otherwise no light would be emitted in the forward direction. The fact that the incident light pulse was efficiently emitted back again into free-space with the same wavelength implies that, a virtual in-plane electric dipole had been created in the Source. Had the dipole's polarization been out-of-plane, no light in essence would have been registered on tracer (1) and subsequently on tracer (3). There is no way that light can propagate in the forward direction, apart from the incident light back-reflection, unless an in-plane dipole near-field emerges within the Source. The incident light's pulse dynamics tailor the virtual-dipole's oscillatory strength and lifetime. The electric dipolar energy, as observed on tracer (2), can be maximized, for example, in response to the incoming light's pulse duration. The ability to register a light signal as a virtual-dipole and recover it back again can be practical for sensing and memory applications. From a different perspective, this interchangeability can open prospects for cavityless stimulated emission.

**Quantum and classical light emitting platforms**

With the elapsed SiV color center in diamond highlights, the TRANSGUIDE's underlying physics should be clear by now. This brings us to a more of an applied discussion, where in what remains of this text, we project the previous findings on selective light emitting platforms.

### Near-infrared light emitters

This section takes us a bit deeper into the near-infrared region, and explores the TRANSGUIDE's potentials with even higher refractive index materials.

Cubic silicon carbide (3C-SiC) has emerged as a promising single photon emitter platform along the telecom range [15]. In Fig. 9a, the transistor-like architecture is applied to a 3C-SiC active medium, where it is demonstrated that ultra-bright directional light emission is realizable in spite of the material's high-refractive index, n ≈ 2.55 [16]. A 50 nm 3C-SiC core is sandwiched in a $SiO_2$ cladding. The light emitting dipoles can be anywhere within the active medium, no restrictions on their locations; the only requirement is parallelism to the Gate. In the bandwidth plot shown in Fig. 9c, the free-space LEE hits a plateau around 92%, with the Purcell factor maintaining values between 1 and 2.

Before going through the next example on semiconductor Quantum Dots (QDs), it makes sense to first identify the relevant state of the art in this regard; namely the GaAs planar antenna system from [8]. It should be noted that the reported ~86% LEE, utilizing $SiO_2$ intermediate layers, considers emission through a semi-infinite glass collection medium, with the refractive index of GaAs reading n = 3.539. When switching to free-space emission, it starts to get tricky, and the antenna's free-space LEE drops to 80%. With that in mind; on the other hand, when applying the TRANSGUIDE to the aforementioned system, considering the configuration shown in Fig. 9b, different behaviors come into play. The corresponding emitter position dependency plot shown in Fig. 9d demonstrates how, the free-space LEE can hit 91% towards the center of the Source, accompanied with Purcell factor values on the order of 1.

The reported LEE values, whether for 3C-SiC or GaAs, consider free-space emission. In an appropriate material packaging, the LEE is likely to pick up additional percents. With these figures in mind, one can already vision the enhanced brightness and radiation profile that await LEDs.

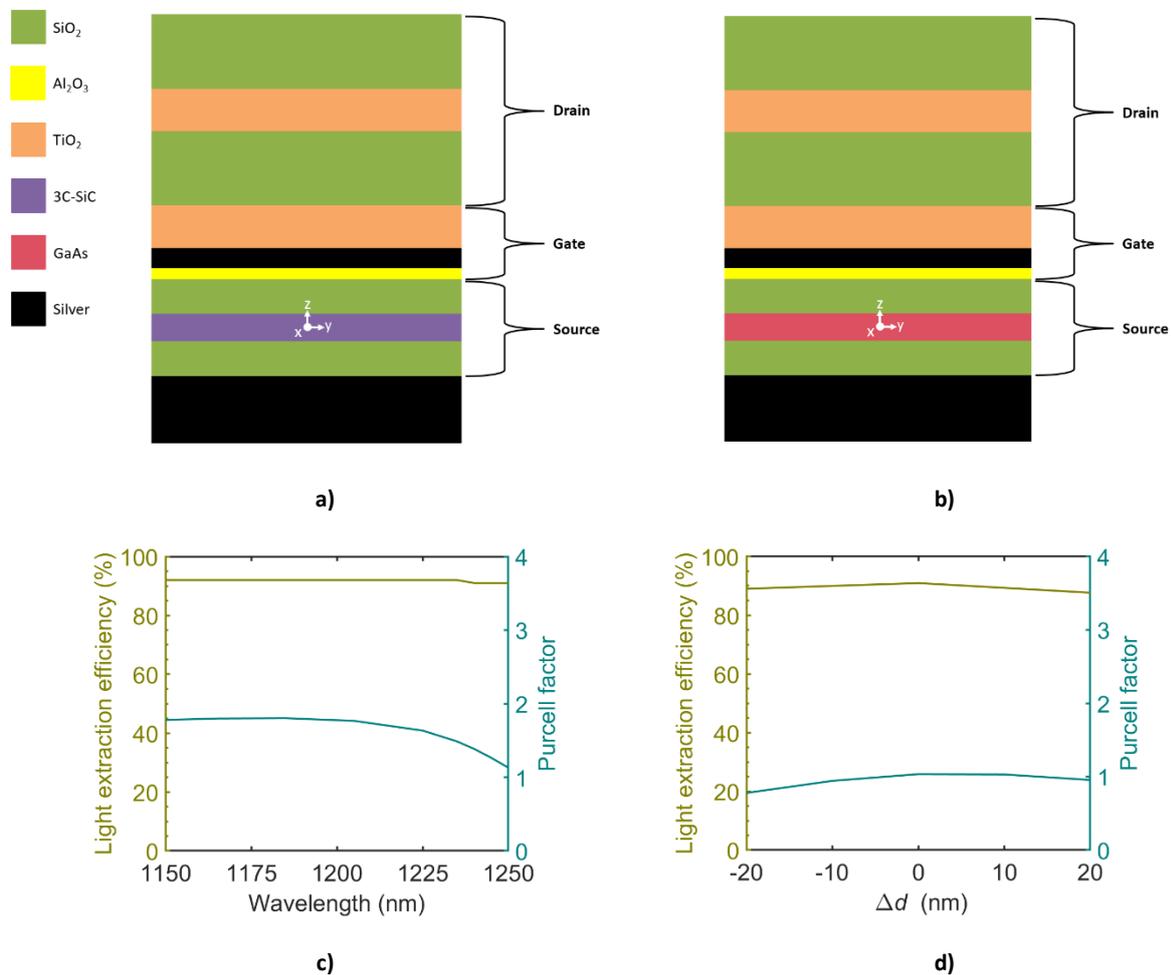

**Fig. 9 Near-infrared emitters (free-space emission).** a) A 3C-SiC thin film in a real-Source trenched-Gate configuration; dimensions: [Forward-bias: (100 nm)], [Source: (114 nm, 50 nm, 114 nm)], [Gate: (14 nm, 25 nm, 115 nm)], [Drain: (200 nm, 111 nm, 200 nm)]. b) A QD embedded in a GaAs thin film, in a real-Source trenched-Gate configuration; dimensions: [Forward-bias: (100 nm)], [Source: (30 nm, 50 nm, 30 nm)], [Gate: (11 nm, 35 nm, 81 nm)], [Drain: (149 nm, 80 nm, 149 nm)]. c) The free-space LEE and the Purcell factor as a function of the emission wavelength. d) The free-space LEE and the Purcell factor as a function of the emitter position displacement from the center of the Source. The QD emits at $\lambda$ = 950 nm.

**Low-refractive index active media**

So far, all previous highlights have been on high-refractive index light emitting platforms. The TRANSGUIDE also holds promises for low-refractive index material-based applications, such as biosensing and Organic Light Emitting Diodes (OLEDs). Hereunder, I give an idea of what you can expect in this regard.

Consider the dibenzoterrylene (DBT) molecules system reported in [10]. The corresponding authors conclude that their planar antenna approach is strongly limited by SPP losses; that having been said, it should not come as a surprise that their antenna's free-space LEE is no more than 60%. By reconfiguring their antenna system into a TRANSGUIDE, SPP losses will no longer be a concern.

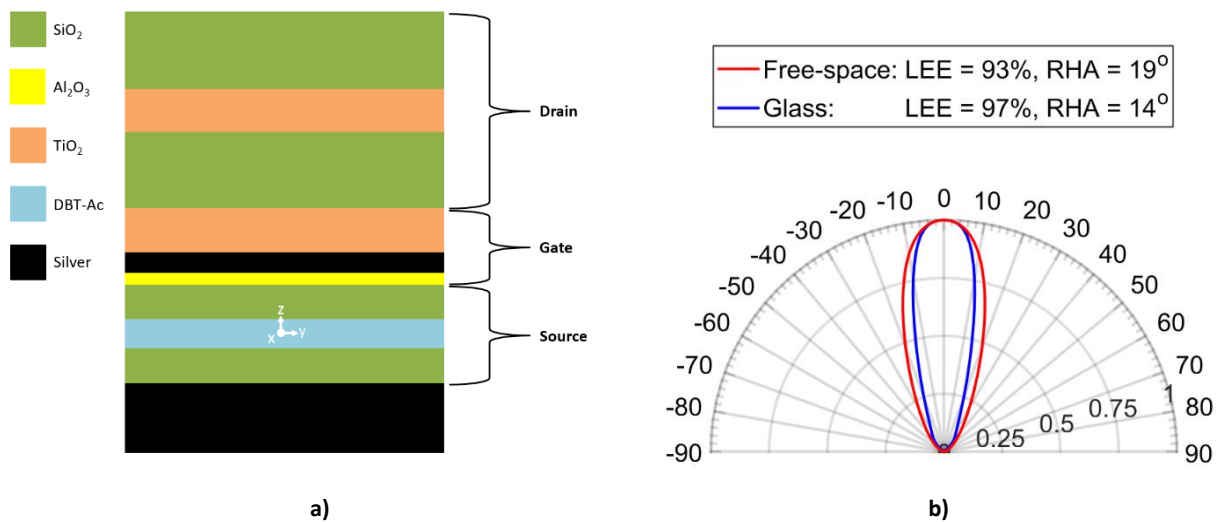

**Fig. 10 DBT molecules emitting at 785 nm**. a) DBT-Ac layer in a real-Source trenched-Gate configuration; dimensions: [Forward-bias: (100 nm)], [Source: (78 nm, 50 nm, 78 nm)], [Gate: (17 nm, 30 nm, 68 nm)], [Drain: (120 nm, 63 nm, 120 nm)]. These dimensions apply explicitly to free-space emission. b) Normalized radiation pattern polar plot, free-space vs glass (n = 1.5) background. Legend: corresponding light extraction efficiency and radiation half angle.

The TRANSGUIDE configuration shown in Fig. 10a, takes into account the same system parameters as in [10]: namely, an n = 1.5 anthracene (Ac) layer containing DBT molecules emitting at 785 nm. With the molecules' dipole orientation parallel to the Gate, the LEE exceeds 93% for free-space emission, and 97% for emission through glass. The radiated power is funneled into a single-lobe, shown in Fig. 10b, with the RHA on the order of 19° and 14°, respectively.

Overall, we are already very close to a 100% LEE, and we should be able to come even closer.

**Conclusion**

Whether the light emitting platform is a low or high-refractive index material, irrespective of the operational wavelength, light extraction efficiencies approaching 100% in conjunction with highly directional emission profiles are realizable in a completely flat ultra-thin structure made of simple materials. It is all about transferring light, sideways, between different waveguide potentials, and pinching. Light emission and absorption are interchangeable processes, and reciprocal systems capable of handling light efficiently in both directions hold promises for numerous photonic applications.

**Acknowledgment**

H. Galal would like to thank Ahmed Samir and Mohammed Nouh.

**Conflict of interest**

EP20020071 patent pending filed on 14.02.2020, entitled "Device for ultra-bright directional light emission". H. Galal is also known in the patent directory with the surname Elmitwalli.